# THE IMPORTANCE OF INCREASING RETURNS TO SCALE IN THE PROCESS OF AGGLOMERATION IN PORTUGAL: A NON LINEAR EMPIRICAL ANALYSIS

**Vitor João Pereira Domingues Martinho**

Unidade de I&D do Instituto Politécnico de Viseu
Av. Cor. José Maria Vale de Andrade
Campus Politécnico
3504 - 510 Viseu
**(PORTUGAL)**
**e-mail:** vdmartinho@esav.ipv.pt

**ABSTRACT**

With this work we try to analyse the agglomeration process in the Portuguese regions, using the New Economic Geography models. In these models the base idea is that where has increasing returns to scale in the manufactured industry and low transport costs, there is agglomeration. Of referring, as summary conclusion, that with this work the existence of increasing returns to scale and low transport cost, in the Portuguese regions, was proven and as such the existence of agglomeration in Portugal.

## 1. INTRODUCTION

With this study we mainly aimed to analyze the process of agglomeration across regions (NUTS II and NUTS III) of Portugal, using non linear models of New Economic Geography, in particular, developments considered by (1-2) Martinho (2004 and 2011) (3)Krugman (1991), (4)Thomas (1997), (5)Hanson (1998) and (6)Fujita et al. (2000). We will also try to compare the results obtained by the empirical models developed by each of these authors.

Although the agglomeration process have appeared more associated with economic geography, it is however noted that it is based, as the polarization, the earlier ideas of (7)Myrdal (1957) and (8)Hirschman (1958), pioneers of the processes of regional growth with characteristics cumulative. The work developed at the level of economic geography, traditional and recent attempt to explain the location of economic activities based on spatial factors. The liberal economic policies, international economic integration and technological progress have created, however, new challenges that promote agglomeration (9)(Jovanovic, 2000). So, have been developed new tools for economic geography, such as increasing returns, productive linkages, the multiple equilibria (with the centripetal forces in favor of agglomeration and centrifugal against agglomeration) and imperfect competition. These contributions have allowed some innovations in modeling the processes of agglomeration, which has become treatable by economists, a large number of issues. In particular the inclusion of increasing returns in the analytical models, which led to the call of increasing returns revolution in economics (Fujita et al., 2000). (10-12)Krugman (1994, 1995 and 1998) has been the central figure in these developments. (13)Fujita (1988),(14) Fujita et al. (1996) and (15)Venables (1996), in turn, have been leaders in the development and exploration of the implications of economic models of location, based on increasing returns. These developments have helped to explain the clustering and "clustering" of companies and industries.

Hanson, in 1998, taking into account the model of Krugman (1991) and the extent of Thomas (1997) this model, had a good theoretical and empirical contribution to empirically examine, with reduced forms, the relationship between increasing returns to scale, costs transportation and geographical concentration of economic activity.

## 2. THE MODEL

The model of Krugman (1991) describes himself, then, as follows:

$$Y_i = (1-\mu)\phi_i + \mu\lambda_i w_i, \text{ income equation (1)}$$

$$G_i = \left[\sum_j \lambda_j (w_j e^{\tau d_{ij}})^{1-\sigma}\right]^{\frac{1}{1-\sigma}}, \forall i, \text{ índex price equation (2)}$$

$$w_i = \left[\sum_j Y_j (G_j e^{-\tau d_{ij}})^{\sigma-1}\right]^{\frac{1}{\sigma}}, \forall i, \text{ nominal wage equation (3)}$$

$$\frac{w_i}{w_j} = \left(\frac{G_i}{G_j}\right)^{\mu}, \text{ real wage equation (3)}$$

In these equations, Yi is the income in region i, wi the wage in region i, $\phi_i$ is the percentage of agricultural workers in region i, Gi the price index for manufactured goods in the region i and dij is the distance between each pair of locations. In equilibrium the region i share $\lambda_i$ employed in sector of manufactured goods which is equal to the fraction of companies located in manufactured goods in region i, ni/n. Alternatively Thomas (1997) presents the following extension of the model of Krugman (1991):

$$Y_i = \lambda_i L w_i, \forall i, \text{ income equation (5)}$$

$$P_i H_i = (1-\mu) Y_i, \forall i, \text{ housing price equation (6)}$$

$$G_i = \left[\sum_j \lambda_j (w_j e^{\tau d_{ij}})^{1-\sigma}\right]^{\frac{1}{1-\sigma}}, \forall i, \text{ price índex equation (7)}$$

$$w_i = \left[\sum_j Y_j (G_j e^{-\tau d_{ij}})^{\sigma-1}\right]^{\frac{1}{\sigma}}, \forall i \text{ nominal wage equation (8)}$$

$$\frac{w_i}{P_i^{1-\mu} G_i^{\mu}} = \frac{w_j}{P_j^{1-\mu} G_j^{\mu}}, \forall i \neq j, \text{ real wage equation (9)}$$

Yi is the income in region i, wi the wage in region i, L the total supply of workers for the manufactured goods sector, $\lambda_i$ the percentage of employees in the sector of manufactured products, Pi the price of housing in region i, the Gi price index for manufactured goods in region i, Hi the supply of housing in the region i and dij is the distance between each pair of locations.

Recently Fujita et al. (2000) also presented an alternative model:

$$Y_i = \mu \lambda_i w_i + (1-\mu)\phi_i, \text{ income equation (10)}$$

$$G_i = \left[\sum_j \lambda_j (w_j T_{ji})^{1-\sigma}\right]^{1/(1-\sigma)}, \text{ price equation (11)}$$

$$w_i = \left[\sum_j Y_j T_{ij}^{1-\sigma} G_j^{\sigma-1}\right]^{1/\sigma}, \text{ nominal wage equation (12)}$$

$$\omega_i = w_i G_i^{-\mu}, \text{ real wage equation (13)}$$

Yi is the income in region i, wi the wage in region i, $\phi_i$ is the percentage of agricultural workers in the region i, $\lambda_i$ the percentage of employees in the sector of manufactured products, Gi price index for manufactured goods in region i, and Tij transport costs between regions i and j.

The parameters to be estimated, these models are $\sigma$ the elasticity of substitution between manufactured goods, $\mu$ the share of expenditure on manufactured goods and $\tau$ the transport costs to send a unit of manufactured goods in a unit distance.

Note that, as can be seen, the three models are very similar, the main difference is that Thomas (1997) have considered building housing sector (power anti-agglomeration) and have created more than one equation and Fujita et al . (2000) have considered transport costs as variables and not considered as parameters in their models Krugman (1991) and Thomas (1997).

It should be noted also that the equations of the income of the previous models, it is assumed that agricultural workers earn the same wage everywhere, given that agricultural goods are freely transported. Were chosen, on the other hand, units such that there are $\mu$ workers in manufacturing and $1-\mu$ agricultural workers.

It could be argued that as industrial workers who are potential users, then locations with large concentrations also tend to have high demand for manufactured goods. This concentration of consumers and producers to some extent explains the cumulative process that may lead to agglomeration phenomena.

Following procedures of Hanson (1998), substituting equations (1) and (4) in (2) yields the reduced equation (14), substituting equations (5), (6) and (9) in (8) obtain the reduced equation (15) and substituting equations (10) and (13) in (11) yields the reduced equation 16, namely:

$$\log(w_i) = C + \sigma^{-1} \log\left(\sum_j Y_j w_j^{\frac{\sigma-1}{\mu}} e^{-\tau(\sigma-1)d_{ij}}\right) + v_i, \quad (14)$$

$$\log(w_i) = D + \sigma^{-1} \log\left(\sum_j Y_j^{\frac{\sigma(\mu-1)+1}{\mu}} H_j^{\frac{(1-\mu)(\sigma-1)}{\mu}} w_j^{\frac{\sigma-1}{\mu}} e^{-\tau(\sigma-1)d_{ij}}\right) + \eta_i, \quad (15)$$

$$\log(w_i) = F + \sigma^{-1} \log\left(\sum_j Y_j w_j^{\frac{\sigma-1}{\mu}} T_{ij}^{-(\sigma-1)}\right) + \psi_i, \quad (16)$$

Thus Hanson (1998a) solved the problem of lack of price indices for manufactured products and prices for housing at more disaggregated geographic levels. In the last two equations C, D and F are constants and parameters, and $\eta_i$, $v_i$ and $\psi_i$ are error terms.

Furthermore, if the sources of correlation are unobservable factors that are constant over time, then these factors can be controlled using a specification with differentiation in time, which makes the variables expressed in growth rates. Given the dearth of statistical data for the Portuguese regions and the small size of the Portuguese territory, this third alternative to solve the problems of endogeneity seems to be the most viable and as such will be adopted in this work.

Using the differences in the timing of the regression equations, the equation (14) becomes:

$$\Delta \log(w_{it}) = \sigma^{-1} \begin{bmatrix} \log(\sum_j Y_{jt} w_{jt}^{\frac{\sigma-1}{\mu}} e^{-\tau(\sigma-1)d_{ij}}) - \\ \log(\sum_j Y_{jt-1} w_{jt-1}^{\frac{\sigma-1}{\mu}} e^{-\tau(\sigma-1)d_{ij}}) \end{bmatrix} + \Delta v_{it}, \quad (17)$$

Equation (15) is also:

$$\Delta \log(w_{it}) = \sigma^{-1} \begin{bmatrix} \log(\sum_j Y_{jt}^{\frac{\sigma(\mu-1)+1}{\mu}} H_{jt}^{\frac{(1-\mu)(\sigma-1)}{\mu}} w_{jt}^{\frac{\sigma-1}{\mu}} e^{-\tau(\sigma-1)d_{ij}}) - \\ \log(\sum_j Y_{jt-1}^{\frac{\sigma(\mu-1)+1}{\mu}} H_{jt-1}^{\frac{(1-\mu)(\sigma-1)}{\mu}} w_{jt-1}^{\frac{\sigma-1}{\mu}} e^{-\tau(\sigma-1)d_{ij}}) \end{bmatrix} + \Delta \eta_{it}, \quad (18)$$

Similarly, equation (16) is as follows:

$$\Delta \log(w_{it}) = \sigma^{-1} \begin{bmatrix} \log(\sum_j Y_{jt} w_{jt}^{\frac{\sigma-1}{\mu}} T_{ijt}^{-(\sigma-1)}) - \\ \log(\sum_j Y_{jt-1} w_{jt-1}^{\frac{\sigma-1}{\mu}} T_{ijt-1}^{-(\sigma-1)}) \end{bmatrix} + \Delta \psi_{it} \quad (19)$$

On balance, taking into account the developments of the New Economic Geography, a value $\sigma/(\sigma-1)$ greater than one indicates that the production is subject to increasing returns to scale. This is because, for the New Economic Geography economies of scale arise through the number of varieties of manufactured goods will be greater the lower the elasticity of substitution $\sigma$. Thus, the lower the elasticity of substitution is further away from one the value of $\sigma/(\sigma-1)$ and the greater the increasing returns to scale.

(16)Krugman (1992) shows that if $\sigma(1-\mu)>1$, then increasing returns to scale are sufficiently weak or the fraction of the manufactured goods sector is sufficiently low and the range of possible equilibria depends on transport costs. If $\sigma(1-\mu)<1$, then increasing returns are sufficiently strong or the fraction is sufficiently high, such as economic activity is concentrated geographically to any value $\tau$.

### 3. THE DATA USED

Considering the variables of the model presented previously, and the availability of statistical information, we used the following data at regional level: temporal data from 1987 to 1994 for the five regions (NUTS II) in mainland Portugal and for the various manufacturing industries existing in these regions, from the regional database of Eurostat statistics (Eurostat Regio of Statistics 2000), and data for the period 1995 to 1999, for the five regions and for total manufacturing, from the INE (National Accounts 2003).

### 4. ESTIMATIONS MADE

Analysis of the results presented in Table 1, obtained in the estimations for the period 1987 to 1994, it appears that these are slightly different for the reduced equations of the three models considered, with the estimates made with the equation of the Thomas model present statistically better results. Possibly because it is an equation to work harder and thus beyond the centripetal forces of agglomeration processes favorable to consider also the centrifugal forces of anti-agglomeration by immobile factors. Anyway, the point that it confirms the results obtained with the estimates of three equations of some importance, but small, transport costs, given the low values of the parameter $\tau$. Looking at the increasing returns to scale, calculating, as noted, the value $\sigma/(\sigma-1)$, it appears that this is always greater than one, reflecting the fact that there were increasing returns in the Portuguese regions in this period. It should be noted also that the parameter values $\mu$ are unreasonably high in all three estimations, however, as stated (17)Head et al. (2003) there is a tendency for these values fall around the unit in most empirical work.

According to Table 2, with the results obtained in the estimations for the period 1995 to 1999, there is again that these are slightly different, although the estimation results with the model equation of Thomas (with agricultural employment as a force anti- agglomeration) are again more satisfying, submitted by the parameter values $\mu$ to less than unity as would be expected in view of economic theory. Note that when considering the stock of housing as centrifugal force, although the results show evidence of greater economies of scale (as noted by the data analysis, or had a close relationship between this variable and nominal wages) are statistically less satisfactory. There is also that $\sigma/(\sigma-1)$ values are always higher than unity, is confirmed also for this period the existence of increasing returns to scale, although with a moderate size, given the value $\sigma(1-\mu)$, i.e. 1.830, in the model Thomas. Since as noted above, when $\sigma(1-\mu)>1$ increasing returns to scale are sufficiently weak or the fraction of the manufactured goods sector is sufficiently low and the range of possible equilibria depends on the costs of transportation. Should be noted that the parameter $\tau$ is not statistical significance in Krugman model and present a very low value in the model of Thomas, a sign that transportation costs have left the already small importance that had in the previous period, which is understandable given the improvements in infrastructure that have been check in Portugal, mainly through the bulk of the structural supports that have come to our country after the appointed time our entry into EEC (European Economic Community), within a set of programs financed by various funds, including Cohesion Fund, among others.

**Table 1:** Results of estimations of the models of Krugman, Thomas and Fujita et al., in temporal differences, for the period 1987-1994, with panel data (at NUTS II level)

**Krugman Model in differences**

$$\Delta \log(w_{it}) = \sigma^{-1} \left[ \log(\sum_j Y_{jt} w_{jt}^{\frac{\sigma-1}{\mu}} e^{-\tau(\sigma-1)d_{ij}}) - \log(\sum_j Y_{jt-1} w_{jt-1}^{\frac{\sigma-1}{\mu}} e^{-\tau(\sigma-1)d_{ij}}) \right] + \Delta v_{it}$$

| Parameters and $R^2$ | Values obtained |
|---|---|
| $\sigma$ | 5.110* (3.611) |
| $\mu$ | 1.262* (6.583) |
| $\tau$ | 0.862* (1.622) |
| $R^2$ | 0.111 |
| DW | 1.943 |
| SEE | 0.196 |
| Nº observations | 284 |
| $\sigma/(\sigma-1)$ | 1.243 |

**Thomas Model in differences**

$$\Delta \log(w_{it}) = \sigma^{-1} \left[ \begin{array}{l} \log(\sum_j Y_{jt}^{\frac{\sigma(\mu-1)+1}{\mu}} H_{jt}^{\frac{(1-\mu)(\sigma-1)}{\mu}} w_{jt}^{\frac{\sigma-1}{\mu}} e^{-\tau(\sigma-1)d_{ij}}) - \\ \log(\sum_j Y_{jt-1}^{\frac{\sigma(\mu-1)+1}{\mu}} H_{jt-1}^{\frac{(1-\mu)(\sigma-1)}{\mu}} w_{jt-1}^{\frac{\sigma-1}{\mu}} e^{-\tau(\sigma-1)d_{ij}}) \end{array} \right] + \Delta \eta_{it}$$

| Parameters and $R^2$ | Values obtained |
|---|---|
| $\sigma$ | 9.076* (2.552) |
| $\mu$ | 1.272* (21.181) |
| $\tau$ | 0.713* (2.053) |
| $R^2$ | 0.145 |
| DW | 1.932 |
| SEE | 0.192 |
| Nº Observações | 284 |
| $\sigma/(\sigma-1)$ | 1.124 |

**Fujita et al. Model in differences**

$$\Delta \log(w_{it}) = \sigma^{-1} \left[ \begin{array}{l} \log(\sum_j Y_{jt} w_{jt}^{\frac{\sigma-1}{\mu}} T_{ijt}^{-(\sigma-1)}) - \\ \log(\sum_j Y_{jt-1} w_{jt-1}^{\frac{\sigma-1}{\mu}} T_{ijt-1}^{-(\sigma-1)}) \end{array} \right] + \Delta \psi_{it}$$

| Parameters and $R^2$ | Values obtained |
|---|---|
| $\sigma$ | 2.410* (31.706) |
| $\mu$ | 1.612* (3.178) |
| $R^2$ | 0.111 |
| DW | 1.990 |
| SEE | 0.215 |
| Nº Observações | 302 |
| $\sigma/(\sigma-1)$ | 1.709 |

**Note: Figures in brackets represent the t-statistic. * Coefficients statistically significant to 5%. ** Coefficient statistically significant 10%.**

**Table 2:** Results of estimations of the models of Krugman, Thomas and Fujita et al., in temporal differences, for the period 1995-1999, with panel data (the level of NUTS III)

| Krugman Model in differences | |
|---|---|
| $$\Delta \log(w_{it}) = \sigma^{-1} \left[ \log(\sum_j Y_{jt} w_{jt}^{\frac{\sigma-1}{\mu}} e^{-\tau(\sigma-1)d_{ij}}) - \log(\sum_j Y_{jt-1} w_{jt-1}^{\frac{\sigma-1}{\mu}} e^{-\tau(\sigma-1)d_{ij}}) \right] + \Delta v_{it}$$ | |
| **Parameters and $R^2$** | **Values obtained** |
| $\sigma$ | 7.399** (1.914) |
| $\mu$ | 1.158* (15.579) |
| $\tau$ | 0.003 (0.218) |
| $R^2$ | 0.199 |
| DW | 2.576 |
| SEE | 0.023 |
| Nº observations | 112 |
| $\sigma/(\sigma-1)$ | 1.156 |
| **Thomas Model in differences (with agricultural workers to the H)** | |
| $$\Delta \log(w_{it}) = \sigma^{-1} \left[ \log(\sum_j Y_{jt}^{\frac{\sigma(\mu-1)+1}{\mu}} H_{jt}^{\frac{(1-\mu)(\sigma-1)}{\mu}} w_{jt}^{\frac{\sigma-1}{\mu}} e^{-\tau(\sigma-1)d_{ij}}) - \log(\sum_j Y_{jt-1}^{\frac{\sigma(\mu-1)+1}{\mu}} H_{jt-1}^{\frac{(1-\mu)(\sigma-1)}{\mu}} w_{jt-1}^{\frac{\sigma-1}{\mu}} e^{-\tau(\sigma-1)d_{ij}}) \right] + \Delta \eta_{it}$$ | |
| **Parameters and $R^2$** | **Values obtained** |
| $\sigma$ | 18.668* (3.329) |
| $\mu$ | 0.902* (106.881) |
| $\tau$ | 0.061* (2.383) |
| $R^2$ | 0.201 |
| DW | 2.483 |
| SEE | 0.023 |
| Nº observations | 112 |
| $\sigma/(\sigma-1)$ | 1.057 |
| $\sigma(1-\mu)$ | 1.830 |
| **Thomas Model in differences (with housing stock to the H)** | |
| $$\Delta \log(w_{it}) = \sigma^{-1} \left[ \log(\sum_j Y_{jt}^{\frac{\sigma(\mu-1)+1}{\mu}} H_{jt}^{\frac{(1-\mu)(\sigma-1)}{\mu}} w_{jt}^{\frac{\sigma-1}{\mu}} e^{-\tau(\sigma-1)d_{ij}}) - \log(\sum_j Y_{jt-1}^{\frac{\sigma(\mu-1)+1}{\mu}} H_{jt-1}^{\frac{(1-\mu)(\sigma-1)}{\mu}} w_{jt-1}^{\frac{\sigma-1}{\mu}} e^{-\tau(\sigma-1)d_{ij}}) \right] + \Delta \eta_{it}$$ | |
| **Parameters and $R^2$** | **Values obtained** |
| $\sigma$ | 11.770 (1.205) |
| $\mu$ | 1.221* (8.993) |
| $\tau$ | 0.003 (0.314) |
| $R^2$ | 0.173 |
| DW | 2.535 |
| SEE | 0.024 |
| Nº observations | 112 |
| **Fujita et al. Model in differences** | |
| $\Delta \log(w_{it}) = \sigma^{-1} \left[ \log(\sum_j Y_{jt} w_{jt}^{\frac{\sigma-1}{\mu}} T_{ijt}^{-(\sigma-1)}) - \log(\sum_j Y_{jt-1} w_{jt-1}^{\frac{\sigma-1}{\mu}} T_{ijt-1}^{-(\sigma-1)}) \right] + \Delta \psi_{it}$ | |
| **Parameters and $R^2$** | **Values obtained** |
| $\sigma$ | 5.482* (4.399) |
| $\mu$ | 1.159* (14.741) |
| $R^2$ | 0.177 |
| DW | 2.594 |
| SEE | 0.023 |
| Nº observations | 112 |
| $\sigma/(\sigma-1)$ | 1.223 |

**Note: Figures in brackets represent the t-statistic. * Coefficients significant to 5%. ** Coefficients significant acct for 10%.**

### 5. CONCLUSIONS

In light of what has been said above, we can conclude the existence of agglomeration processes in Portugal (around Lisboa e Vale do Tejo) in the period 1987 to 1999, given the transport costs are low and it was shown by $\sigma/(\sigma-1)$ and the $\sigma(1-\mu)$ values obtained in the estimations made with the reduced forms of the models presented above, there are increasing returns to scale in manufacturing in the Portuguese regions. This is because, according to the New Economic Geography, in a situation with low transport costs and increasing returns to scale, productive linkages can create a circular logic of agglomeration, with links "backward" and "forward". What makes the producers are located close to their suppliers (the forces of supply) and consumers (demand forces) and vice versa. The driver of the process is the difference in real wages, i.e., locations that, for some reason, have higher real wages attract more workers (which are also potential consumers), calls "forward" which, in turn, attract more companies to meet the requirements of demand, calls "backward." With a greater concentration of companies in the same location, the products are shifted to lower distances, saving on transport costs and, as such, prices may be lower, nominal wages may be higher and so on. On the other hand, when certain factors are real estate (land), they act as centrifugal forces that oppose the centripetal forces of agglomeration. The result of the interaction between these two forces, traces the evolution of the spatial structure of the economy.

Note that the results obtained with the estimates of Thomas model equations are statistically more satisfactory, possibly because they consider these equations in addition to the centrifugal forces present in increasing returns, also by centrifugal forces, in this work, the number of employees in the sector agricultural. It should be noted, finally, that transport costs have had some importance in the evolution of the space economy in Portugal, which amount has been decreasing in recent years, which is understandable given the investments that have been made in terms of infrastructure structures, especially after the appointed time our entry into the European Economic Community in 1986, with the support that has been under structural policies.